\documentclass[onecolumn,nofootinbib]{revtex4}

\usepackage{setspace}
\usepackage{amsmath,amssymb}
\usepackage{graphicx}
\usepackage{dcolumn}
\usepackage{bm,color}
\usepackage{accents}
\usepackage{amssymb,float}
\usepackage{amsmath}
\usepackage{comment}
\usepackage{multirow}
\usepackage{tikz}
\usepackage{enumerate}

\newcolumntype{P}[1]{>{\centering\arraybackslash}p{#1}}
\newcolumntype{M}[1]{>{\centering\arraybackslash}m{#1}}
\usepackage{hyperref}
\hypersetup{
    colorlinks=true,
    linkcolor=red,
    filecolor=magenta,      
    citecolor=blue
}
\newcommand{\bea}{\begin{eqnarray}}
\newcommand{\eea}{\end{eqnarray}}
\newcommand{\be}{\begin{equation}}
\newcommand{\ee}{\end{equation}}

\newcommand{\dd}{\mathrm{d}}
\begin{document}

\title{Gravitational Collapse in pure Gauss-Bonnet gravity}

\author{Konstantinos F. Dialektopoulos}
\email{kdialekt@gmail.com}
\affiliation{Department of Mathematics and Computer Science, Transilvania University of Brasov, 500091 Brasov, Romania}
\affiliation{Laboratory of Physics, Faculty of Engineering, Aristotle University of Thessaloniki, 54124 Thessaloniki, Greece}
\affiliation{Department of Physics, Nazarbayev University, 53 Kabanbay Batyr avenue, 010000 Astana, Kazakhstan}

\author{Daniele Malafarina}
\email{daniele.malafarina@nu.edu.kz}
\affiliation{Department of Physics, Nazarbayev University, 53 Kabanbay Batyr avenue, 010000 Astana, Kazakhstan}

\author{Naresh Dadhich}
\email{nkd@iucaa.in}
\affiliation{Inter-University Centre for Astronomy \& Astrophysics, Post Bag 4, Pune, 411 007, India}

\date{\today}

\begin{abstract}
We study the process of gravitational collapse in pure Gauss-Bonnet gravity. In the homogeneous dust collapse, we show that the $D=7$ pure Gauss-Bonnet theory has gravitational dynamics indistinguishable from Einstein's theory in $D=4$, meaning that collapsing particle feel the same potential as in the classical 4-dimensional general relativistic case. In $D<7$ pure Gauss-Bonnet gravity becomes weaker, while in $D>7$ it becomes stronger, with respect to General Relativity. %However, this is not the case for bound/unbound collapse, where pure Gauss-Bonnet is always weaker/stronger than General Relativity.
In the inhomogeneous dust collapse we find the mass modes in the expansion of the energy density in any dimensions that lead to either naked singularities or black holes as final states of collapse. 
\end{abstract}

\maketitle

\section{Introduction}\label{sec:intro}

The singularity theorems state that under certain conditions, if trapped surfaces form during collapse, then a singularity is inevitable \cite{Penrose:1964wq,Hawking:1970zqf,Hawking:1976ra,Senovilla:1998oua}.
One of the assumptions necessary for the validity of the theorems is that General Relativity (GR) holds during collapse. However, the appearance of singularities is usually taken as an indication that GR does not hold beyond certain density scale and the theory must be replaced by a new, yet unknown, theory of gravity. This has naturally led to researchers to investigate gravitational collapse and the validity of similar theorems in theories alternative to GR \cite{Hawking:1972qk,Sotiriou:2011dz,Brown:2018hym}.

Gauss-Bonnet theory (GB) is a higher-order theory belonging to Lovelock's class \cite{Lovelock:1971yv}. It appears as the $N=2$ order in the Lovelock action, with $N=1$ being the Ricci scalar in GR. 
GB has been studied for many years as one of the most interesting higher-order theories of gravity \cite{Dadhich:2013bya,Chakraborty:2016qbw,Pons:2014oya}, 
which still yields second order equations despite the action being polynomial in Riemann, and it also arises in heterotic string theory as the low energy limit correction to the Einstein-Hilbert action \cite{Bento:1995qc,1987NuPhB.291...41G}. Lovelock's theorem \cite{Lovelock:1972vz} states that terms of order $N$ become non-trivial at $D>2N$, where $D$ is the dimensionality of spacetime. Hence, the Gauss-Bonnet term is a topological invariant in $D = 4$ and contributes in the dynamics only for $D>4$. However, if one couples it to a scalar field (or an arbitrary function of a scalar field) \cite{Antoniou:2017acq,Bakopoulos:2018nui,Doneva:2022ewd,Guo:2020sdu}, not only it becomes dynamical, but also it gives an interesting property, the so-called scalarization of black holes, which means that the no-hair theorem is violated and there appear scalar-hair in black hole solutions. For this reason several extensions of GR that include the Gauss-Bonnet term either non-linearly in 4-D or coupled to other fields have been studied in the past \cite{Bajardi:2019zzs,Bahamonde:2019swy,Bahamonde:2022chq,Bahamonde:2022lvh,Odintsov:2020vjb,Odintsov:2022zrj,Oikonomou:2021kql,Oikonomou:2020oil,Dialektopoulos:2022kiv}.

Additionally, it has been recently suggested \cite{Glavan:2019inb} that there may exist a non-trivial 4-D limit of Einstein-Gauss-Bonnet (EGB) gravity in a way that could allow to avoid Lovelock's theorem. Specifically, if one multiplies the GB term in $D>4$ Einstein-Gauss-Bonnet gravity, with the term $1/(D-4)$, and then takes the limit $D \rightarrow 4$, it may be possible to get a non-trivial class of theories in $D=4$ that propagate only two degrees of freedom. Unfortunately, the theory in \cite{Glavan:2019inb} is possessed by strong coupling both in cosmology and around black hole solutions. A new attempt without the strong coupling has been pursued in \cite{Bonifacio:2020vbk}, but the number of degrees of freedom propagated becomes 2+1, meaning that the theory can be considered as a subclass of Horndeski gravity, where there's an extra scalar propagating degree of freedom. In \cite{Aoki:2020lig} the authors suggest that if one violates the 4-D diffeomorphism invariance, one gets a consistent (meaning without strong coupling), but non-covariant $D\rightarrow 4$ EGB gravity. Since then, many applications have been investigated in the context of this theory
\cite{Lu:2020iav,Hennigar:2020lsl,Malafarina:2020pvl,Shaymatov:2020yte}.

The above discussion suggests that there is value in exploring gravitation theories in $D>4$ and their connection to Einstein's theory. The term \textit{pure Lovelock gravity} has been used to describe theories that consider only one $N$th order term of the Lovelock polynomial, without summing over the lower orders. In this sense, pure GB gravity will utilize the $N=2$ Lovelock term, which is the GB term without assuming a summation with the Ricci scalar, namely ignoring the Einstein-Hilbert part of the action. One  of the distinguishing characteristics of the pure Lovelock theory is that, in $D = 2N+1$ (i.e. $D=5$ for GB) gravity is kinematic \cite{Camanho:2015hea,Dadhich:2012cv,Dadhich:2015ivt}, which means that the Lovelock Riemann (namely the generalization of Riemann to Lovelock order) tensor can be given entirely in terms of the corresponding Ricci. In addition, as shown in Sec.~\ref{sec:pure-GB-gravity}, in the absence of a cosmological constant, the vacuum solution is trivial in $D = 5$. Non-trivial black holes exist only in $D \geq 6$ in pure GB gravity. Other applications in the context of the theory can be found in \cite{Mukherjee:2020lld,Dadhich:2021vdd,Mukherjee:2021erg,Abdujabbarov:2015rqa}.

In this paper, we consider gravitational collapse of inhomogeneous dust in pure GB and investigate how it relates to the corresponding solutions in GR.
Gravitational collapse of inhomogeneous dust in GR is known to produce naked singularities under some conditions \cite{Newman:1985gt,Waugh:1988ud,Shapiro:1991zza,Dwivedi:1992fh,Joshi:1993zg}. 
These conditions relate to the `strength' of gravity, the initial density profile and the initial velocity profile for the infalling matter. Many investigations have been carried out in GR to determine the influence of density and velocity profiles on the final outcome of collapse \cite{Dwivedi:1994qs,Joshi:1998su,2002PhRvD..65j1501J,Harada:2001nj,Mena:2001dr,Joshi:2004tb,Joshi:2011rlc}. 
In the present work on the other hand we focus on the strength of gravity and thus consider collapse in pure GB theory in arbitrary dimensions and investigate how it compares to corresponding models in GR in 4-dimension as well as GR in higher dimensions \cite{Banerjee:2002sy,Ghosh:2001fb,Goswami:2002he,Ghosh:2002ky,Ghosh:2002mf,Goswami:2004gy,Goswami:2006ph}. Gravitational collapse of dust in pure GB and EGB has been considered in \cite{Maeda:2005ci,Maeda:2006pm,Maeda:2007uu,Jhingan:2010zz}. 

The article is organized as follows: In Sec.~\ref{sec:pure-GB-gravity} we describe the theory and derive its equations of motion. As expected, since it is a sub-class of Lovelock gravity, the equations of motion will be of second order. Also, in Sec.~\ref{sec:pure-GB-gravity} we consider a static and spherically symmetric metric and find vacuum solutions in arbitrary $D-$dimensions in the presence of a cosmological constant, $\Lambda$. We show that, in $D = 5$ pure GB gravity does not have any non-trivial vacuum solutions; in $D = 6$ it is weaker than GR; in $D = 7$ it behaves exactly like GR, while in $D \geq 8$ it becomes stronger, where we understand the terminology `weaker' and `stronger' as relating to the strength of the gravitational potential $1/r^\alpha$ compared to GR. In Sec.~\ref{sec:gravitational-collapse} we study gravitational collapse, in particular, we consider homogeneous dust collapse in spacetimes with negative, zero and positive spatial curvature. 
%Interestingly enough, even though in the marginally bound case (i.e. zero curvature), GB collapse behaves as described before, in the bound case (i.e. positive curvature), the singularity always forms later with respect to the corresponding case in GR, while in the unbound case (i.e. negative curvature), it always forms earlier. 
Moreover, we consider marginally bound inhomogeneous dust collapse and we find the critical mass modes, i.e. the coefficients in the radial expansion of the mass profile, in $D \geq 6$ where both naked singularities and black holes are formed. In GR the first two modes always allow for the occurrence of naked singularities, the third one is the critical mode, where both naked singularities and black holes can form depending on some condition, while for all higher modes we only have the formation of black holes. Similarly, we find that in $D = 6$ pure GB gravity, where as we discussed gravitational interactions are weaker than GR, the critical mode is the fifth, while $D = 7$ behaves like GR. In $D = 8$, there is no critical mode as the first and second can always lead to the formation of a naked singularity while the third or higher always leads to a black hole. In $D=9$ the second mode separates between the two outcomes while for $D>9$ only the first mode may allow for the occurrence of naked singularities. A brief discussion of the results is then presented in Sec.~\ref{conc}.
For simplicity, throughout the paper we use natural units setting $G=c=1$.
%second mode becomes the critical one, while in $D = 9$ even from the first mode we can have the formation of black holes. 

\section{Gauss-Bonnet gravity and black hole solutions}\label{sec:pure-GB-gravity}

Lovelock gravity is the most general metric theory of gravity that gives second order field equations in an arbitrary number of spacetime dimensions $D$. In particular, its Lagrangian density is a polynomial that reads
\begin{equation}\label{eq:Lovelock-action}
    \mathcal{L} = \sum _{N=0}^{\bar{N}} c_N \mathcal{L}_N\,,
\end{equation}
where
\begin{equation}
    \mathcal{L}_N = \frac{1}{2^N}\delta _{\alpha _1\beta _1...\alpha _N\beta _N}^{\mu _1\nu _1...\mu _N \nu _N}\prod _{r = 1}^N R^{\alpha _r \beta _r}{}_{\mu _r \nu _r}\,,
\end{equation}
with 
\begin{equation}
    \delta _{\alpha _1\beta _1...\alpha _i\beta _i}^{\mu _1\nu _1...\mu _N \nu _N} = (2N)! \delta _{[\alpha _1} ^{\mu_1} \delta _{\beta _1} ^{\nu _1} ... \delta _{\alpha _N} ^{\mu_N} \delta _{\beta _N]} ^{\nu _N} \,,
\end{equation}
being the generalized Kronecker delta and $R^{\alpha\beta}{}_{\mu\nu}$ the Riemann tensor. Only those terms with $N < D/2$ contribute in the dynamics of the theory and thus the dimensionality in \eqref{eq:Lovelock-action} can be taken to be $D = 2 N +2$ for even and $D = 2N +1$ for odd dimensions. In that way, in $D = 3$ and $ D = 4$, it coincides with GR, but in higher dimensions in contains more terms in the action.

If one considers only the $N$-th order terms in the Lovelock action, the theory is called pure Lovelock gravity. For $N = 2$ we have the pure Gauss-Bonnet gravity, whose action in $D$ dimensions reads
\begin{equation}
    \mathcal{S} = \int \dd ^D x \sqrt{-g} \left(\mathcal{G} - \Lambda \right) + \mathcal{S}_{\rm matter}\,,
\end{equation}
where $\mathcal{G}=R^2-4R_{\mu\nu}R^{\mu\nu}+R_{\mu\nu\rho\sigma}R^{\mu\nu\rho\sigma}$ is the Gauss-Bonnet term and $\Lambda$ is the cosmological constant.

Varying the action with respect to the metric we get
\begin{equation}\label{eq:metric_eq_GB5}
    H_{\mu\nu}+\Lambda g_{\mu\nu} = T_{\mu\nu}\,,
\end{equation}
where $H_{\mu\nu}$ is the Lanczos tensor given by
\begin{equation}
    H_{\mu\nu} = 2 \left[ R _{\mu\nu} - 2 R_{\mu\alpha}R^{\alpha}{}_{\nu} -2 R^{\alpha\beta} R_{\mu\alpha\nu\beta} +R_{\mu}{}^{\alpha\beta\gamma}R_{\nu\alpha\beta\gamma}\right] - \frac{1}{2}g_{\mu\nu} \mathcal{G}\,,
\end{equation}
and the energy-momentum tensor of the matter fields obtained from the variation of $\mathcal{S}_{\rm matter}$ is considered to be that of a perfect fluid
\begin{equation}
    T_{\mu\nu} = \left(\rho + p\right) u_{\mu}u_{\nu} + p g_{\mu\nu}\,,
\end{equation}
with $u_{\mu} = \delta ^0{}_\mu$ the fluid's 4-velocity, $\rho(t,r)$ the energy density and $p(t,r)$ the isotropic pressure of the matter source.

%\section{Black hole solutions}\label{sec:black-holes-solutions}

In order to describe a black hole solution let us consider the following static and spherically symmetric line element
\begin{equation} 
    \dd s^2 = - A(r) \dd t^2 + B(r) \dd r^2 + r^2 \dd \Omega ^2_{D-2}\,.
\end{equation}
The $(tt)$ and $(rr)$ field equations in vacuum in $D$ dimensions are written as
\begin{gather} \label{11}
 2C r (1-B(r)) B'(r)-C (D-5) B(r) \left(B(r)^2+1\right)+2C (D-5) B(r)^2+\Lambda  r^4 B(r)^3 = 0 \,,\\ \label{12}
 2C r (B(r)-1) A'(r)+A(r) \left(2C (D-5) B(r)-C  (D-5) \left(B(r)^2+1\right)+\Lambda  r^4 B(r)^2\right) = 0 \,,
\end{gather}
where 
\begin{equation}\label{C}
C = \frac{(D-2)(D-3)(D-4)}{2} \;.
\end{equation}
Multiplying Eq.~\eqref{12} by $B/A$ and adding it to the Eq.~\eqref{11}, we get
\begin{equation}
    B = \frac{c}{A}\,,
\end{equation}
where $c$ is an integration constant and in order for the asymptotics to be Minkowski, we can set it to unity. Then solving Eq.~\eqref{11} for $A$, we get the black hole solution as
\begin{equation} \label{A(r)}
    A(r) = 1- \sqrt{M r^{5-D}+\frac{ \Lambda  r^4}{C}}\,.
\end{equation}

As already mentioned in the introduction, the critical case $D = 2N +1 = 5$ (here we have $N = 2$ in Lovelock gravity), does not have any non-trivial vacuum solutions if the cosmological constant is vanishing. Once we assume that $\Lambda \neq 0$ the theory accepts black hole solutions which are of BTZ type. In $D \geq 6$ the gravitational potential falls off as $r^{-(D-5)/2}$ in the absence of $\Lambda$. This means that in $D = 6$ we expect pure GB theory to be weaker than GR, in $D = 7$ we expect it to behave like GR, while in $D \geq 8$ we expect the theory to be stronger. As we will show in the next section, this is indeed the case as it is realised in gravitational collapse. 

\begin{comment}
In 5D the equations of motion are
\begin{gather}
     \frac{A \left(\Lambda r^3 B^3 + 6B' - 6 BB' \right)}{r^3 B^3} = 0\,,\\
     \Lambda B + \frac{6 A' \left(B-1 \right)}{r^3 AB} = 0\,.
\end{gather}
From the first one we get that
\begin{equation}
    B(r) =  \frac{2 \left(\sqrt{36 + 72 M + 3 \Lambda r^4}-6 \right)}{\Lambda r^4 + 24 M}\,,
\end{equation}
and substituting it into the second one we get
\begin{equation}
    A(r) = c \left(6 + \sqrt{36 + 72 M + 3 \Lambda r^4} \right)\,.
\end{equation}

In $D=6$ the equations become
\begin{gather}
    \frac{A \left(24 B^2 + (\Lambda r^4 - 12) B^3 + 24 r B' - 12 B (1+2r B) \right)}{r^4 B^3} = 0\,,\\
    \frac{A\left(24 B -12 + B^2 (\Lambda r^4 -12) \right)+24 r A' (B-1)}{r^4 AB} = 0\,.
\end{gather}
Multiplying the second one with $A/B$ and adding it to the first one, we get
\begin{equation}
    B = \frac{c}{A}\,,
\end{equation}
and in order for the asymptotics to be Minkowski, we set $c=1$. Then solving the first one for $A$, we get
\begin{align}
    A(r)  = 1\pm \sqrt{\frac{M}{r}+\frac{\Lambda r^4}{60}}\,,
\end{align}
where only the one with the minus is a physical one.

Similarly in 7D and 8D we'll have respectively,
\begin{gather}
    A(r) = B(r)^{-1} = 1\pm \sqrt{\frac{M}{r^2}+\frac{\Lambda r^4}{180}}\,,\\
    A(r) = B(r)^{-1} = 1\pm \sqrt{\frac{M}{r^3}+\frac{\Lambda r^4}{420}}
\end{gather}
\end{comment}

\section{Gravitational collapse}
\label{sec:gravitational-collapse}

Let us consider the Lema\'itre-Tolman-Bondi (LTB) line element in comoving coordinates, that is
\begin{equation}\label{eq:LTB-metric}
    \dd s ^2 =  - \dd t^2 + B(t,r) ^2 \dd r^2 + R(t,r)^2 \dd \Omega ^2 _{D-2}\,,
\end{equation}
with $\dd \Omega ^2 _{D-2}$ being the metric of the unit $(D-2)$-dimensional sphere. 

From the $r$-component of the continuity equation in any dimensions we get $p = p(t)$
while from the $t$-component we get
\begin{equation}\label{eq:cont_eq}
    \dot{\rho}+ (p+\rho )\left(\frac{\dot{B}}{B} + (D-2) \frac{\dot{R}}{R}\right) = 0\,.
\end{equation}

The amount of matter contained within the co-moving shell $r$ at the co-moving time $t$ may be obtained from the definition of the quasi-local mass in co-moving coordinates which generalizes the static case of the mass contained within a spherical volume. This is known as the Misner-Sharp mass and it can be written as
\begin{equation}\label{eq:F-1}
    F(t,r) = \frac{R^{D-1}}{R^{2N}}\left(1-B^{-2}R'^2+\dot{R}^2\right)^N\,,
\end{equation}
with $N=2$. The off-diagonal component of the field equations \eqref{eq:metric_eq_GB5} in any dimension gives
\begin{equation}
    \left[B^2 \left(\dot{R}^2+1\right)-R'^2\right] \left(B \dot{R}'-\dot{B} R'\right) = 0\,.
\end{equation}
From the above, we get two branches for collapse depending on the form of $B(t,r)$. The first one is
\begin{equation}
    B(t,r) = \pm \frac{R'}{\sqrt{\dot{R}^2+1}}\,,
\end{equation}
while the second one is
\begin{equation}\label{eq:B_metric_potential}
    B(t,r) = \frac{R'}{E(r)}\,,
\end{equation}
where $E(r)$ is an integration function related to the initial velocity of the collapsing particles. 
The first branch does not give any massive dynamical collapse solutions since, substituting it into the definition of the Misner-Sharp mass Eq.~\eqref{eq:F-1} it is easy to see that we get $F=0$.
Therefore in what follows, we work with \eqref{eq:B_metric_potential} which resembles the metric potential in GR. The remaining non-trivial field equations are
\begin{gather}\label{eq:rho}
    \rho(t,r) = C\frac{F'}{R^{D-2} R'} - \Lambda\,,\\ \label{eq:p}
    p(t) = - C\frac{\dot{F}}{R^{D-2} \dot{R}}+\Lambda\,,
\end{gather}
where $C$ is the constant in Eq. \eqref{C}
and we have used Eq. \eqref{eq:B_metric_potential} to define the mass function $F$ as
\begin{equation}\label{eq:F}
    F(t,r) = R^{D-5} \left(1-E^2 +\dot{R}^2 \right)^2\,.
\end{equation} 
Eq.~\eqref{eq:F} can be used as the equation of motion to be solved in order to obtain the solution $R$ of the collapse model. In the following we shall focus on pressureless, usually called `dust', models. In addition it will be useful to apply the following rescaling to the area-radius $R$, Misner-Sharp mass $F$ and velocity profile $E$:
\begin{gather}\label{R-rescale}
    R(t,r)=ra(t,r) \,,\\ \label{F-rescale}
    F(t,r)=r^{D-1}m(t,r) \,, \\ \label{E-rescale}
    E(r)=\sqrt{1-b(r)r^2} \,.
\end{gather}

With the above rescaling we have replaced the arbitrary integration function $E(r)$ that appears in Eq.~\eqref{eq:B_metric_potential} with the new function $b(r)$. To have a straightforward interpretation of both quantities we may look at Eq.~\eqref{eq:F}. The initial velocity of the collapsing particles located at the shell $r$ is given by $\dot{R}(t_i,r)$, which depends only on the initial matter distribution, i.e. the initial Misner-Sharp mass function $F(t_i,r)$ and $E(r)$. Therefore we may understand both $E(r)$ and $b(r)$ as providing the initial condition for the velocity of particles at the shell $r$. Since we know that $g_{rr}=B^2$, from Eq.~\eqref{eq:B_metric_potential} we can also interpret the function $E(r)$ as related to the spatial curvature. In fact, as we shall see later, for a homogeneous spacetime we have precisely $E^2=1-kr^2$ with $k$ the curvature.

%\subsection{Specific cases}

\subsection{Homogeneous dust collapse}

Let us first consider a pressureless homogeneous fluid in the absence of a cosmological constant, i.e. we shall set $p = 0 = \Lambda$. The above rescaling in Eqs \eqref{R-rescale}-\eqref{E-rescale} becomes $R(t,r) = r a(t) $, $F(t,r) = r^{D-1}m_0$ and $E(r) = \sqrt{1-k r^2}$,  with $m_0$ and $k$ 
being constants. Then, from Eq.~\eqref{eq:rho}-\eqref{eq:F} and \eqref{eq:cont_eq}, the metric \eqref{eq:LTB-metric} becomes,
\begin{equation}
    \dd s^2 = - \dd t^2 + a ^2 \left( \frac{\dd r^2}{1-kr^2} + r^2 \dd \Omega^2_{D-2} \right)\,,
\end{equation}
where the scale factor $a(t)$ should satisfy the equation of motion
\begin{equation}\label{eq:scale-factor-dust}
    m_0 - a^{D-5}\left(k+\dot{a}^2\right) ^2 = 0\,,
\end{equation}
and the energy density of the fluid is
\begin{equation}
    \rho(t) = C(D-1) \frac{m_0}{a^{D-1}}\,.
\end{equation}

\paragraph{Marginally bound collapse:}

This case is obtained for $k=0$ and it is easy to see that in $D=7$ it is formally equivalent to the equation of motion for dust collapse in GR, which is $\dot{a}^2=\tilde{m}_0/a$, with the identification of $\tilde{m}_0=\sqrt{m_0}$.
In fact Eq.~\eqref{eq:scale-factor-dust} with $k=0$ gives the general solution
\begin{equation}\label{eq:scale-factor-dust-k=0}
    a(t) = \left(\frac{D-1}{4}\right)^{\frac{4}{D-1}} \left( \pm m_0^{1/4} t + a_0\right) ^{\frac{4}{D-1}}\,,    
\end{equation}
where the case with the $+$ describes expansion, while the one with the $-$ describes collapse. There are two remaining solutions which are complex. 
It is important to notice that the solution for $a(t)$ in $D=5$, which is given by $a(t)=1-m_0^{1/4}t$, is not physical since the fact that $\ddot{a}=0$ implies that gravity is not dynamical in $D=5$, a fact that is reflected in the exterior black hole solution \eqref{A(r)}, which in turns implies that the only acceptable solution is the flat vacuum one with $\dot{a}=0$ and $m_0=0$, corresponding to $M=0$ for the exterior metric.
It is immediate to see that in $D=7$ one gets $a\sim t^{2/3}$ as in GR. Normalizing the radial coordinate so that at the initial time $t_i=0$ we have $R(0,r) = r$, i.e. $a(0) = 1$, we get the integration constant $a_0$ as
\begin{equation}
    a_0 = \frac{4}{D-1}\,,
\end{equation}    
which leads to
%\begin{equation}\label{eq:scale-factor-k=0}
%    a(t) = \left(\frac{D-1}{4}\right)^{\frac{4}{D-1}} \left( - m_0^{1/4} t + \frac{4}{D-1}\right) ^{\frac{4}{D-1}}\,.    
%\end{equation}
\begin{equation}\label{eq:scale-factor-k=0}
    a(t) =  \left(1 -\frac{D-1}{4} m_0^{1/4} t \right) ^{\frac{4}{D-1}}\,.    
\end{equation}

%{\bf THERE SEEMS TO BE INCONSISTENCY BETWEEN EQS(28) AND (30) FOR D=5, THE FORMER GIVES FOR $k=0$, $ \dot{a} = const$ WHILE THE LATTER NON CONST? IN FIG 1 WHAT HAVE YOU ASSUMED FOR $m_0$?}

%{\bf (DM: I think, as we discussed last time, that by looking at equation \eqref{A(r)} we have that the only allowed exterior solution must have $M=0$ and therefore in the above general expression the only allowed value for $m_0$ is zero, which gives $a$ constant. But the inconsistency is not between (28) and (30), as they both allow for solution (32). From what I understand, the  restriction to $m_0=0$ comes from imposing matching with the only allowed exterior which has $M=0$.)}

\begin{figure}[!ht]
    \includegraphics[scale=0.4]{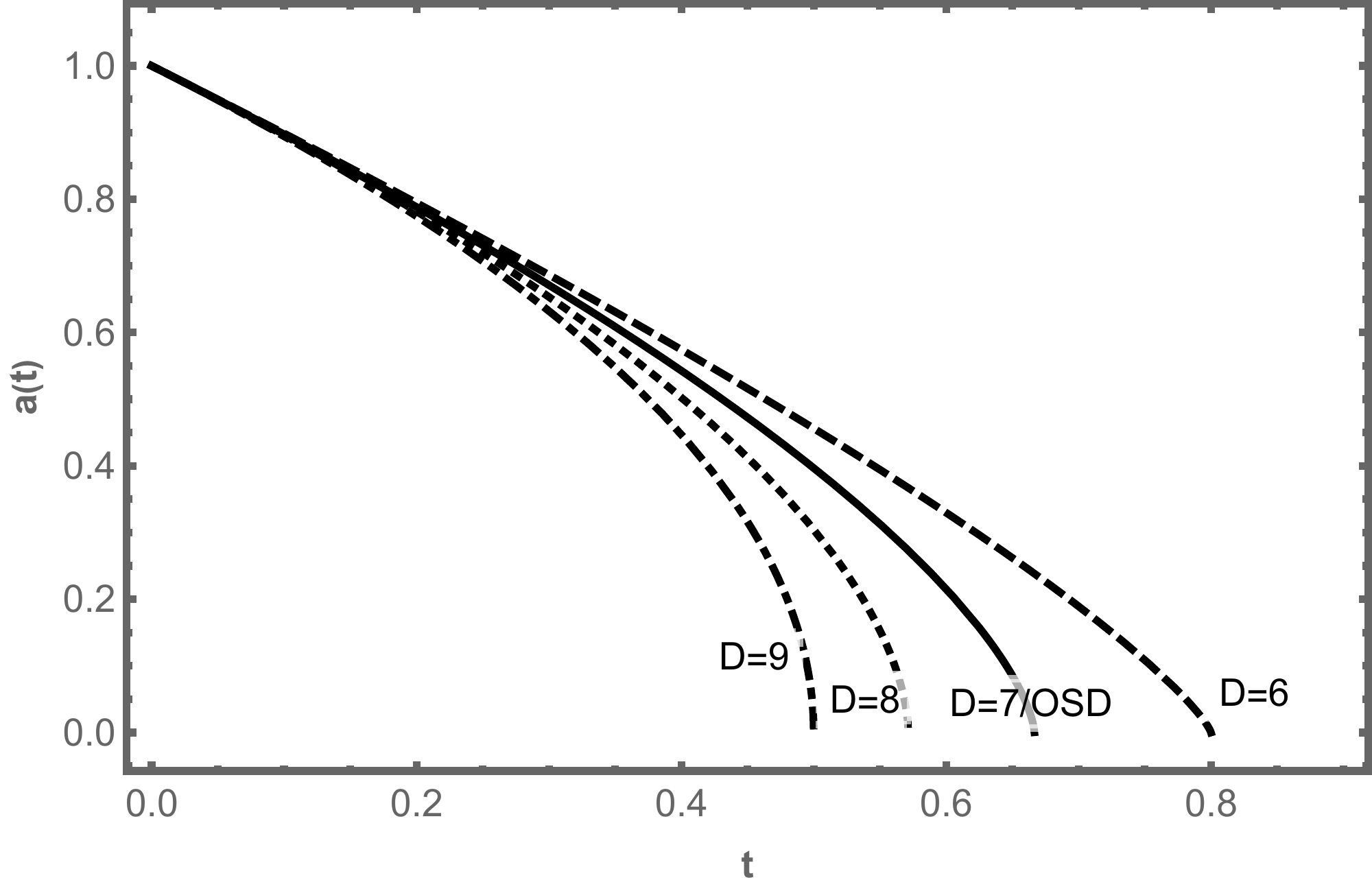}
    \includegraphics[scale=0.4]{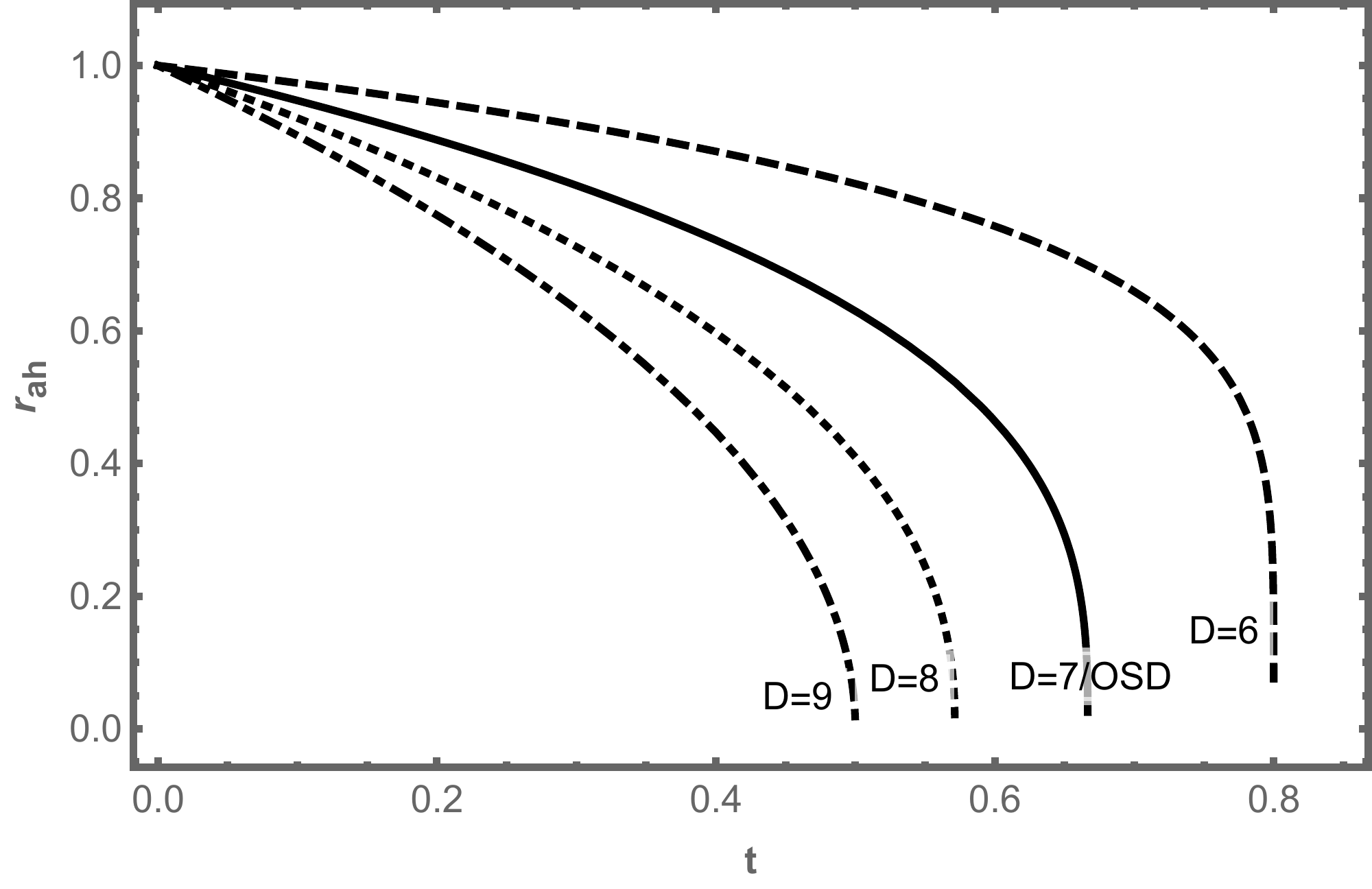}
    \caption{Left panel: The scale factor for homogeneous dust collapse in $D=6,7,8,9$ together with the one in Einstein's gravity (OSD) for $k=0$. We set $m_0=1$ and $a(0) = 1$. Right panel: Radius of the apparent horizon $r_{ah}(t)$ for homogeneous dust collapse in $D=6,7,8,9$ together with the one in Einstein's gravity (OSD) for $k=0$. The OSD case is identical to the $D=7$ in pure Gauss-Bonnet gravity. Since the collapse starts at $t=0$ we see that in $D=6$ each shell $r$ is trapped later than the corresponding shell in GR, while in $D=8$ and $D=9$ this happens faster.}
    \label{fig:dust-flat}
\end{figure}

However, it is obvious that in this case the collapsing fluid cannot be at rest at $t=0$, i.e. $\dot{a}(0) \neq 0$, because otherwise that would imply $m_0 = 0$. This is a consequence of the choice of marginally bound collapse, namely $k=0$. In fact in order to have $\dot{a}(0)\rightarrow 0$ we would have to require $a(0)\rightarrow +\infty$, meaning that the particles can start collapsing from rest if they are initially at spatial infinity, otherwise their initial velocity is always positive. As a consequence the potential of the marginally bound case can be regarded as analogous to the gravitational potential and has been used to study the strength of gravity in the pure Gauss-Bonnet theory \cite{Chakraborty:2016qbw}. In Fig.~\ref{fig:dust-flat} we show the comparison of the scale factor in various dimensions with the Oppenheimer-Snyder-Datt (OSD) case in 4-dimensional GR.
As we can see in the left panel of Fig.~\ref{fig:dust-flat} in $D=6$ (dashed-line) the singularity is delayed compared to general relativity, meaning that gravity is weaker. 
In $D=7$ (solid line) Gauss-Bonnet gravity behaves in the same way as GR, while in $D=8$ (dotted) and $D=9$ (dot-dashed), gravity becomes stronger, thus causing the singularity to occur earlier.

The Kretschmann scalar is given by
\begin{equation}
    K = \frac{12}{a^4} \left[ a^2 \ddot{a}^2 + (k+\dot{a}^2)^2\right]=\frac{3m_0}{a^{D-1}}\frac{(5-D)^2+16}{4} \,.
\end{equation}
%{\bf $K ~ 1/a^4, 1/a^5$ FOR $d=5,6$, HOW DO THE CURVES CROSS IN FIG 2? }
%{\bf (DM: I agree, that looks strange. Also maybe we should remove the case $D=5$ entirely, since only $m_0=0$ is allowed.)}\kos{Indeed, $K \sim 1/a^4, 1/a^5$ in $D = 5, 6$. But $a \sim t, t^{4/5}$ in $D = 5,6$, so in both cases $K \sim 1/t^4$}
In the only physically viable case in $D=5$ the Kretschmann scalar is vanishing, since the scale factor is $a=1$ at all $t$ and $m_0=0$. 
Then, since when $k=0$ in $D=7$ the scale factor has the same behavior as the OSD case, the Kretschmann scalar also exhibits the same behavior, i.e. $K\sim 1/a^6$. The strength of the singularity achieved at $a=0$ then grows with $D$. 
%We also notice that in $D=6$ the curvature singularity forms at the end of the collapse, but it takes more time. 
%\begin{figure}[ht!]
%    \centering
%    \includegraphics[scale=0.5]{Kretschmann_k=0.pdf}
%    \caption{Kretschmann scalar in the GR and in $D=5,6,7$ and $D=8$ pure Gauss-Bonnet gravity with $k=0$. In $D=5$ (solid line) it remains zero. In $D=6$ (short dashed) it diverges later than in $D=7$, which is the same as GR, while in $D=8$ (long dashed) the divergence happens faster.}
%    \label{fig:Kretschmann_k=0}
%\end{figure}
The singularity forms at the time $t_s$ when $a(t_s) = 0$ which in this case is
\begin{equation}
    t_s = \frac{4m_0^{-1/4}}{D-1}\,.
\end{equation}

The formation of trapped surfaces in the collapsing matter cloud is then signalled by the apparent horizon in the interior which is given by the condition $g^{\mu\nu}\partial _{\mu}R \partial _\nu R = 0$, i.e. the surface $R(t,r)$ becomes null. This is expressed by
\begin{equation}\label{rah1}
    1-\sqrt{\frac{F}{R^{D-5}}} = 0 \;\; \text{ or } \;\; 1- \frac{\sqrt{m_0}r^2}{a^{\frac{D-5}{2}}} = 0\,.
\end{equation}
Thus, the apparent horizon curve is 
\begin{equation}\label{eq:r_ah}
    r_{\rm ah} (t) = m_0^{-1/4}a^{\frac{D-5}{4}}\,.
\end{equation}
Obviously $r_{\rm ah} (t)\rightarrow 0$ for $t\rightarrow t_s$ and therefore the apparent horizon forms at the boundary at a time $t<t_s$ and reaches the center at the time of formation of the singularity, thus leaving the singularity hidden at all times.
In the right panel of Fig.~\ref{fig:dust-flat} we show the evolution of the apparent horizon in $D=6,7,8,9$ as well as in the Einstein gravity. As already discussed above, the $D=7$ case coincides with the OSD case, in 5 dimensions there is no dynamics and in $D=6$ the formation of the trapped surfaces as well as the singularity are delayed compared to GR.
%\begin{figure}[!ht]
%    \includegraphics[scale=0.4]{rah_k=0-1.pdf}
%    \caption{Radius of the apparent horizon in the Einstein gravity (OSD), which is identical to the $D=7$ in pure Gauss-Bonnet gravity, for $k=0$. Since the collapse starts at $t=0$ we see that in $D=6$ all the shells are trapped later than in GR. In $D=8$ and $D=9$ this happens faster since GB gravity is stronger than GR.} 
%    \label{fig:rah_k=0}
%\end{figure}

%\begin{figure}[!ht]
%    \centering
%    \includegraphics[scale=0.5]{scale-factor-ratio.pdf}
%    \caption{We show the ratio between the scale factor of pure Gauss Bonnet in $D = 6,7,8$ over the one of OSD collapse in the marginally bound case. The plain line is the $D = 7$ pGB case which coincides with GR, while the long dashed $(D = 6)$ and short dashed $(D = 8)$ lines}
%    \label{fig:scale-factor-ratio}
%\end{figure}

\paragraph{Bound collapse:}

This case is obtained for $k>0$ and in principle one can set $k=1$ with a suitable rescaling of $r$. However, in the plots we have set $m_0=1$ in order to more easily compare with the GR case, which is equivalent to a different rescaling of $r$, and thus we must choose $k<1$. 
Then, as an example, the collapsing solution of Eq.~\eqref{eq:scale-factor-dust} in $D=6$ reads
\begin{equation}
    t(a) = -\frac{4m_0}{k^3} \left(\sqrt{\frac{\sqrt{m_0}}{\sqrt{a}}-k}\right) \, _2F_1\left(\frac{1}{2},3;\frac{3}{2};1-\frac{\sqrt{m_0}}{k\sqrt{a}}\right) \,,
\end{equation}
which needs to be inverted to obtain $a(t)$. 
Here $_2F_1$ is the ordinary hypergeometric function that is given by
\begin{equation}
    _2 F _1 (a,b;c;z) = \sum _{n = 0}^\infty \frac{(a)_n (b)_n}{(c)_n}\frac{z^n}{n!}\,,
\end{equation}
with $(q)_n$ being the Pochhammer symbol defined as
\begin{equation}
    (q)_n =  
     \begin{cases}
  1 & \text{for } \, n = 0 \,, \\
  q (q+1)...(q+n-1) & \text{for }\, n > 0 \,.
  \end{cases}
\end{equation}
In contrast with the $k=0$ case, the bound collapse case allows for the fluid to be at rest at a finite radius at the initial time $t_i=0$.
%The first thing to notice is that with $k\neq 0$ the behavior of collapse departs from the OSD for any $D$, including the case $D=7$. In fact the singularity is delayed if $k=1$ for all $D>5$. This shows that the effect of the curvature (remember that the constant $k$ is related to the initial velocity profile or equivalently to the curvature of space) in pure GB dominates over the role of the gravitational potential. In other words, collapsing particles that start from rest at a finite radius will fall towards the singularity at a slower pace in pure GB with respect to GR regardless of the dimension.
Again we see that the case $D=7$ is formally identical to GR, since Eq.~\eqref{eq:scale-factor-dust} is the same as the one in GR.

\begin{figure}[!ht]
    \centering
    \includegraphics[scale=0.4]{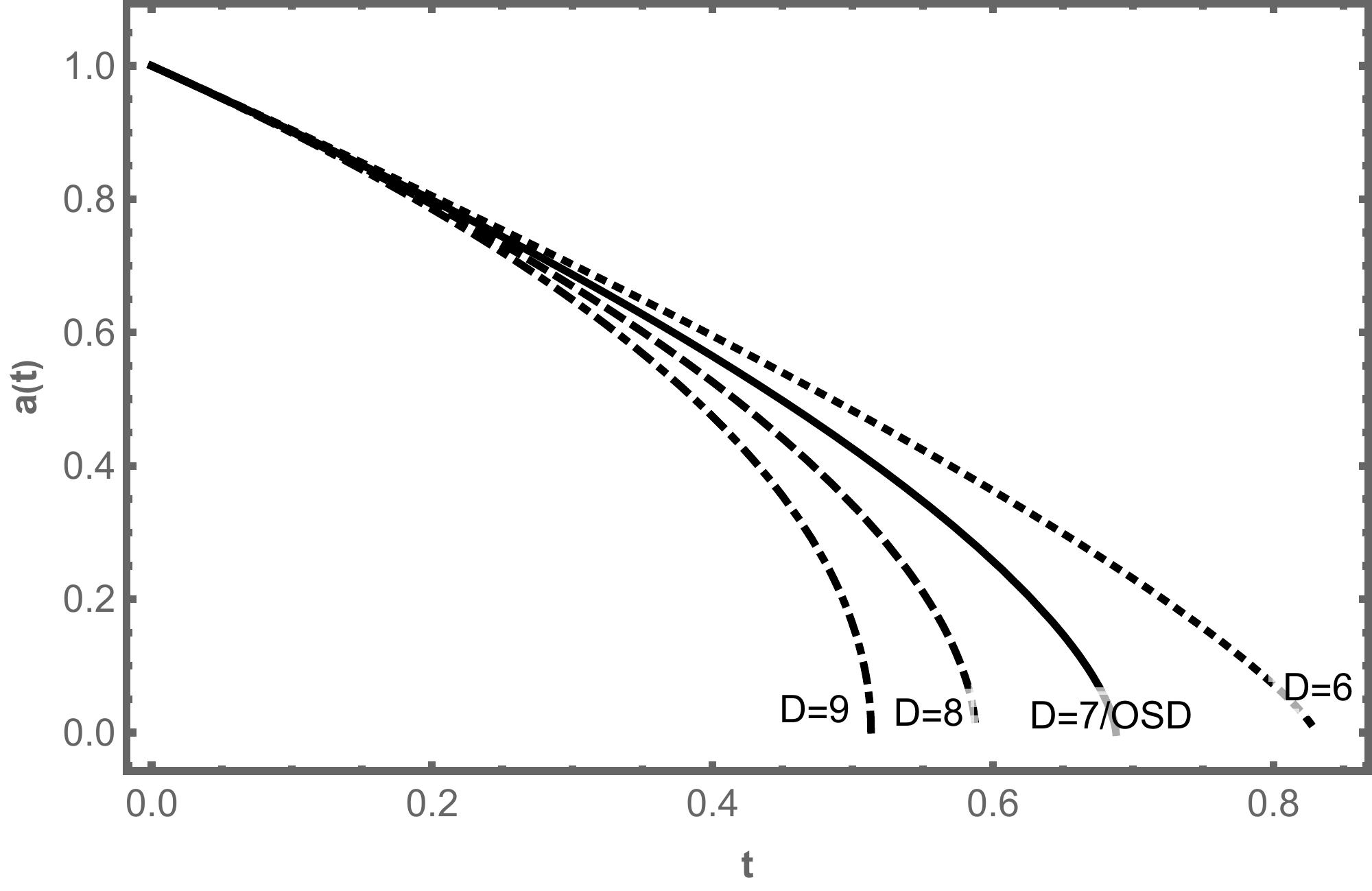}
    \includegraphics[scale=0.4]{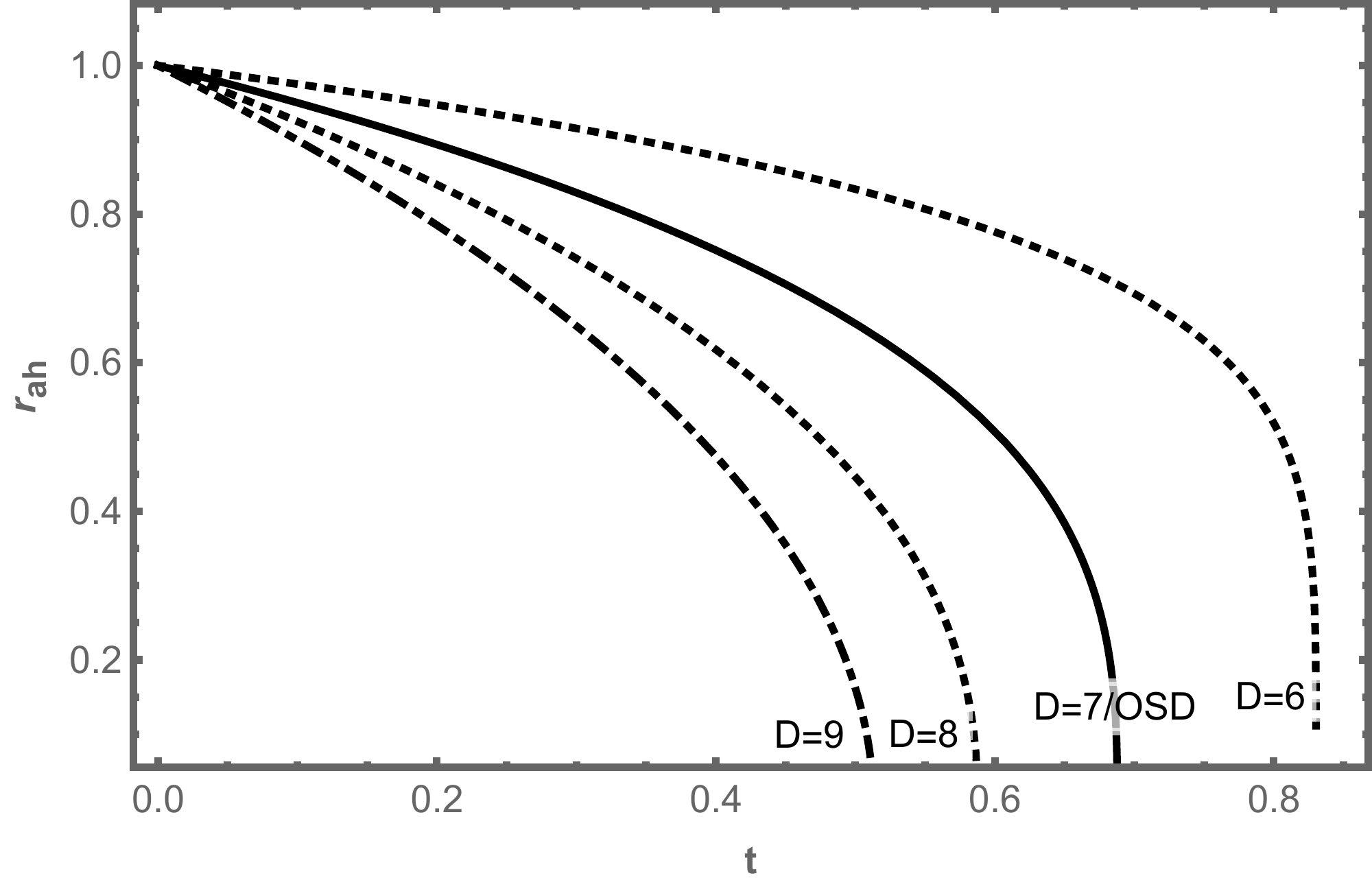}
    \caption{Left panel: We plot the scale factor $a(t)$ in $D=6$ (dotted line), $D=7$ (solid line), $D = 8$ (dashed) and $D=9$ (dot-dashed), together with the OSD case (solid line), for $k=0.1$. We set $m_0=1$ and $a(0) = 1$. Notice than even though we have spatial curvature, $D =7$ pure GB shows the same behaviour as in GR. Right panel: Radius of the apparent horizon in the OSD model together with $D = 7$ pure GB (solid line) and in $D=6$ (dotted line), $D=8$ (dashed) and $D = 9$ (dot-dashed) in pure Gauss-Bonnet gravity, in the case when $k=0.1$.}
    \label{fig:dust-closed}
\end{figure}

In Fig.~\ref{fig:dust-closed} we compare the above solutions for bound pure GB collapse to the OSD case in GR. In the left panel we show the behavior of the scale factor in $D=6,7, 8$ and $9$ as compared to the OSD case, while in the right panel we show the apparent horizon curve. The qualitative behavior is the same as in GR, i.e. the apparent horizon form at the boundary and reaches $r=0$ at the time of formation of the singularity, but its development is delayed with respect to the GR case for $D=6$ and otherwise for $D\geq 8$.

\paragraph{Unbound collapse:}

\begin{figure}[!ht]
    \centering
    \includegraphics[scale=0.4]{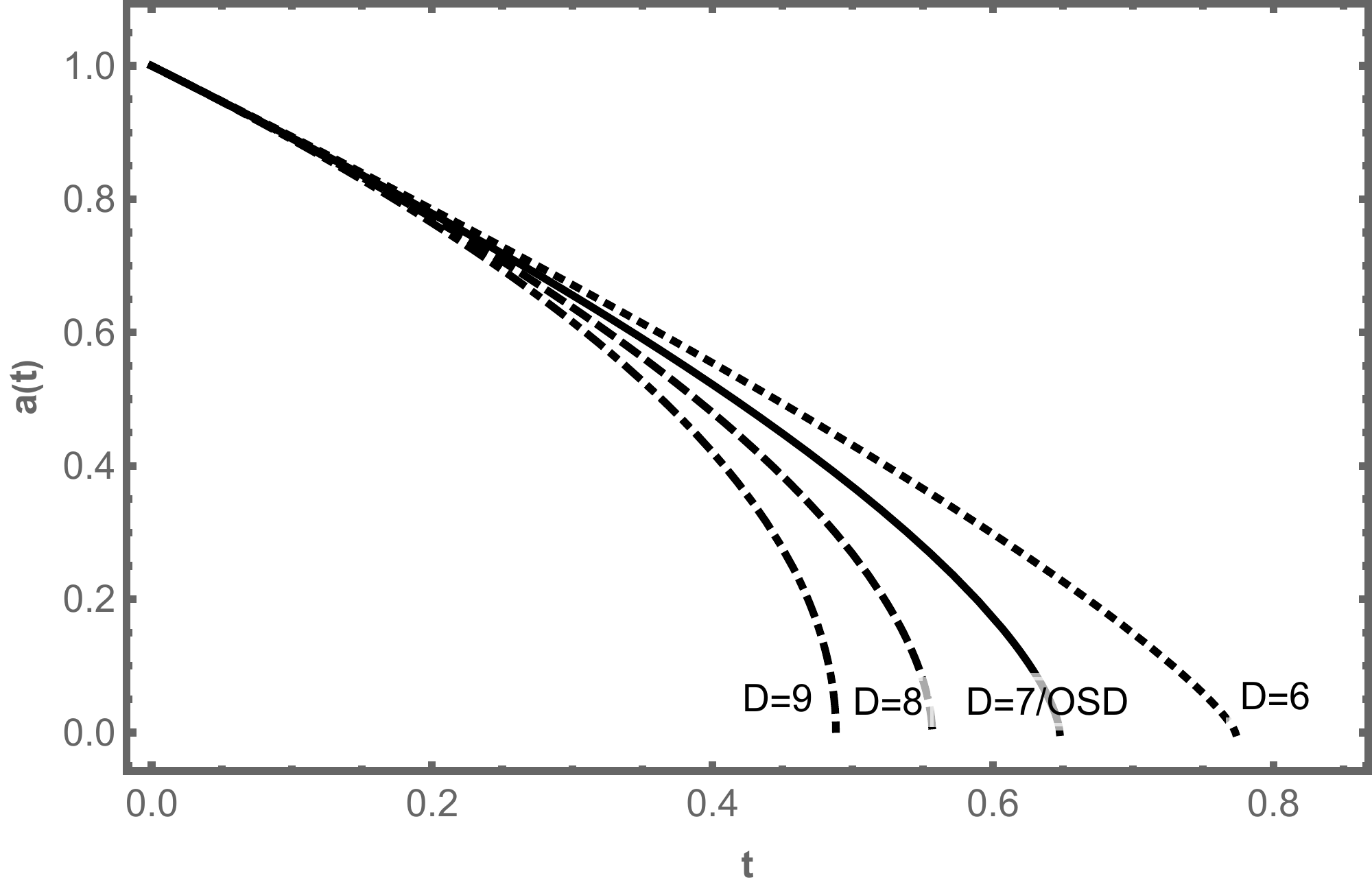}
    \includegraphics[scale=0.4]{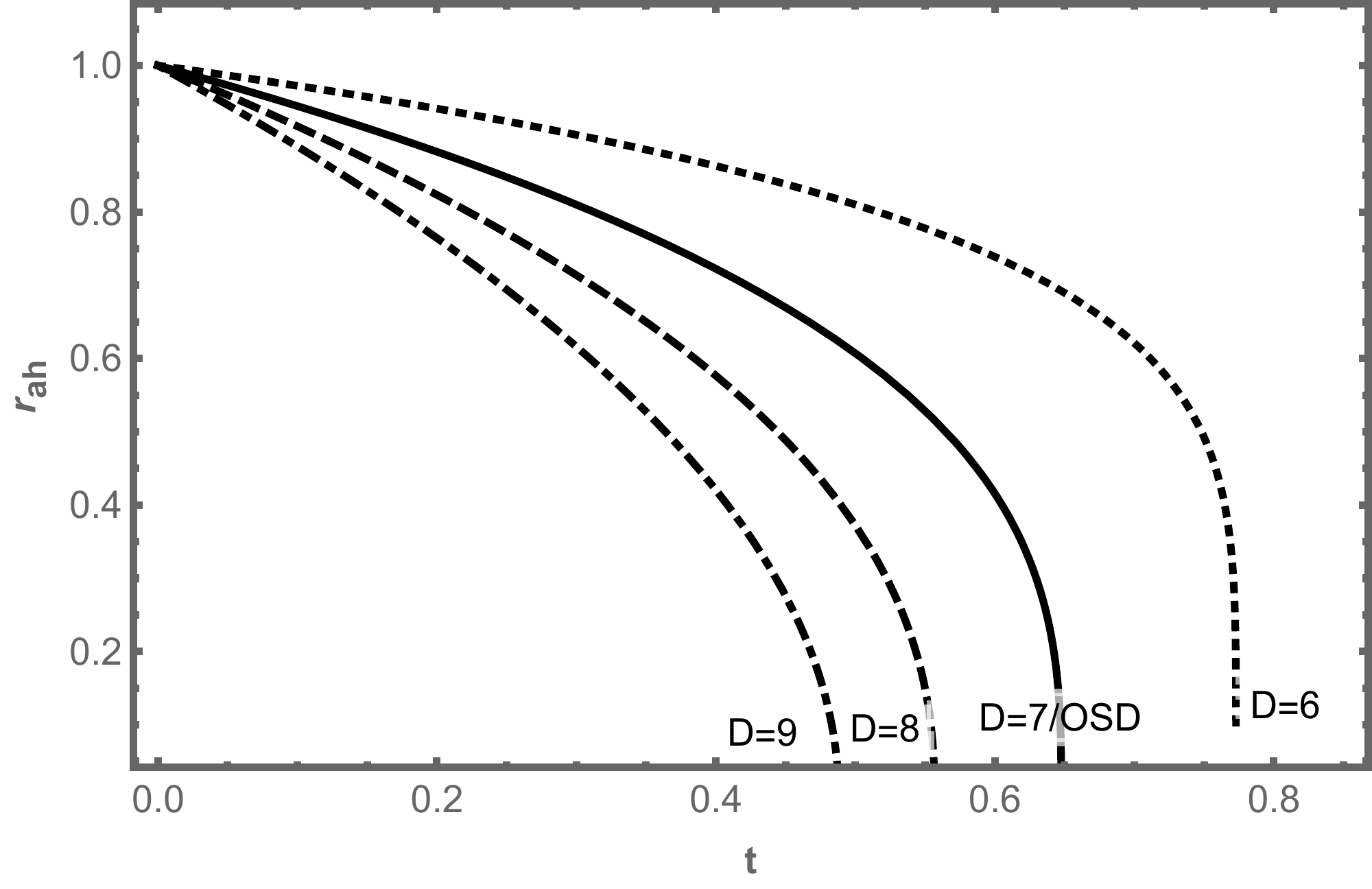}
    \caption{Left panel: The scale factor $a(t)$ in $D=6$ (dotted), $D = 7$ pure Gauss Bonnet, together with the OSD case (solid) and $D = 8$ (dashed), $D = 9$ (dot-dashed) pure GB gravity, for $k = -0.1$. We set $m_0 = 1$ and $a(0) = 1$. Right panel: Radius of the apparent horizon in the OSD case with $k=-1$ (solid line) compared to the $D=6$ (dotted), $D=7$ (solid), $D=8$ (dashed) and $D = 9$ (dot-dashed) pure Gauss-Bonnet gravity, also with $k=-0.1$.}
    \label{fig:dust-open}
\end{figure}

This case is obtained for $k<0$. As an example, the collapsing solutions of Eq.~\eqref{eq:scale-factor-dust} for $k=-1$ in $D=7$ and $D=8$ read respectively
\begin{gather}
    t(a) = t_0-\frac{a}{\sqrt{|k|}}\sqrt{1-\frac{\sqrt{m_0}}{|k| a}}+\frac{\sqrt{m_0}}{|k|^{3/2}}\tanh ^{-1}\left(\sqrt{1-\frac{\sqrt{m_0}}{|k| a}} \right)\,,\\
    t(a) = \, t_0-\frac{a}{\sqrt{-k}} \, _2F_1\left(-\frac{2}{3},\frac{1}{2};\frac{1}{3};-\frac{\sqrt{m_0}}{k a^{3/2}}\right)\,.
\end{gather}
The unbound case corresponds to particles having positive initial velocity at spatial infinity and again we see that in $D=7$ it is formally identical to GR.
%in pure GB such collapse occurs always faster than in GR for all values of $D$. This shows again that the role of curvature in pure GB gravity dominates over the behavior of the potential when comparing with GR.
To illustrate this, in the left panel of Fig.~\ref{fig:dust-open} we compare the solutions for $D=6, 7,8$ and $9$ to the OSD case. 
The apparent horizon is given by Eq.~\eqref{eq:r_ah} and in the right panel of Fig.~\ref{fig:dust-open} we show its evolution over time in $D=6,7,8$ and $9$ in pure GB gravity as compared to GR. 
%As suggested by the previous considerations, trapped surfaces occur faster than in GR for all $D$ in the unbound collapse scenario.

\subsection{Inhomogeneous dust collapse}

Let us now consider the case with $\Lambda = 0$ but allow for the scale factor to have an $r$ dependence, namely describing an inhomogeneous fluid. To implement this we set the scaling as in Eqs.~\eqref{R-rescale}, \eqref{F-rescale} and \eqref{E-rescale} 
%\begin{align}
%    F(t,r ) &= r^{D-1}m(t,r)\,,\\
%    R(t,r) &= r a(t,r) \,,\\
%    E(r) &= \sqrt{1- b(r) r^2}\,,
%\end{align}
so that Eqs.~\eqref{eq:rho}, \eqref{eq:p} and \eqref{eq:F} become
\begin{gather}\label{eq1}
    \rho(t,r) = C\frac{(D-1) m+r m'}{a^{D-2} \left(r a'+a\right)}\,,\\ \label{eq2}
    p(t) = - C\frac{\dot{m}}{\dot{a} a^{D-1}}\,,\\ \label{eq3}
    m(r) = a^{D-5} \left(b(r) + \dot{a}^2\right)^2\,.
\end{gather}
Considering an inhomogeneous dust cloud, meaning $p = 0$, implies that the mass profile is a function of only $r$. Inhomogeneous dust collapse in GR is usually referred to as Lema\'itre-Tolman-Bondi \cite{Lemaitre:1933gd,Tolman:1934za,Bondi:1947fta} and it has been shown to allow for the occurrence of naked singularities \cite{Joshi:1993zg}. Considering an inhomogeneous density profile, with vanishing pressures leads to inhomogeneous mass and velocity profiles and to a scale factor that depends on both $r$ and $t$. Then for the marginally bound case, for which $b(r) = 0$, we get
\begin{equation}\label{eq:scale-factor-k=0-inhomogeneous}
    a(t,r) = \left[1- \frac{D-1}{4}  \left( m(r)^{1/4}\right)t\right]^{\frac{4}{D-1}}\,,
\end{equation}
where, similarly to the homogeneous case, we have set the initial condition as $a(0,r) = 1$. Notice that if $m(r)$ is a constant $m_0$ the solution \eqref{eq:scale-factor-k=0-inhomogeneous} coincides with \eqref{eq:scale-factor-k=0}. Both in GR and in pure GB to obtain the solution of the inhomogeneous case we just need to replace $m_0$ with $m(r)$. Notice that again the case $D=7$ is equivalent to the GR case if we identify the density profile $\tilde{m}$ in GR with the density profile in pure GB via $\tilde{m}(r)=\sqrt{m(r)}$. Each co-moving radius $r$ becomes singular along the curve $t_s(r)$ corresponding to the time when the scale factor vanishes, i.e.
\begin{equation}
    t_s (r)= t_i + \frac{4}{(D-1)m(r)^{1/4}}\,,
\end{equation}
where $t_i=0$ is the initial time from which the collapse develops. The apparent horizon develops for each shell $r$ at the time $ t_{ah}(r)$ which is given by the implicit solution of $g^{\mu\nu}\partial _{\mu}R \partial _\nu R = 0$. In our case, this condition is again given by Eq.~\eqref{rah1}, or
$ R(t_{ah}(r),r)^{D-5} = F(r)$, or equivalently, making use of the scaling,
\begin{equation}
     r^4 m(r) = a(t_{ah}(r),r)^{D-5}\,.
\end{equation}
Using Eq.~\eqref{eq:scale-factor-k=0-inhomogeneous} we find the apparent horizon curve as
\begin{equation}
    t_{ah}(r) = t_s(r) - \frac{4}{D-1}r^\frac{D-1}{D-5}m(r)^{\frac{1}{D-5}} \,.
\end{equation}
%{\bf (DM: I am not putting everything in bold as I will be adding quite a bit from here onwards.)}

It is straightforward to notice that $t_{ah}(0)=t_s(0)=t_0$ for all $D>5$, since  $(D-1)/(D-5)> 1$ and $(D-1)/(D-5)\rightarrow 1$ for $D$ growing. 
%Therefore $t_{ah}(0)=t_s(0)$ and the nature of the singularity is determined by the first non vanishing term in the expansion of $t_c(r)$.
A generic inhomogeneous mass profile can be expanded near $r=0$ as
\begin{equation}
    m(r) = m_0 + \frac{m_n}{n} r^n +o(r^n)\,, 
    %\quad \text{with } m_n <0\,,
\end{equation}
with $n\geq 1$ and where $m_0$ is related to the initial density at the center via $C(D-1)m_0 =\rho(0,0)= \rho_i(0)$. Here we have defined the $n$-th mass mode $m_n/n$ in order to simplify the notation, so that in the following we have $m'=m_nr^{n-1}$. 
To ensure that the density profile decreases outwards one must impose $m_n <0$ for the first non vanishing mass mode. This also ensures that throughout collapse shell crossing singularities do not occur. These are `weak' singularities that appear when shells overlap and looking at Eq.~\eqref{eq1} we see that the condition for the appearance of shell crossing singularities is $R'=a+ra'=0$. Avoidance of shell crossing singularities is guaranteed if $a+ra'>0$ throughout collapse\cite{Hellaby:1985zz}. Notice also that in general for a physically realistic profile one would also want to impose that the density does not present a cusp at the center, thus requiring $m_1=0$.
We may expand $t_s(r)$ near $r=0$ as
\begin{equation}
t_s(r)=\frac{4}{(D-1)m_0^{1/4}}\left(1-\frac{m_nr^n}{4m_0}+...\right)\,,
\end{equation}
Under the above conditions we then see that the singularity curve $t_s(r)$  originates at the center at the time $t_0$ and moves to outer radii at later times, at least in a close neighborhood of $r=0$. On the other hand for $m_n>0$ we would get a density profile increasing radially outwards, which is less physically realistic, and the singularity curve reaching the center at later times with respect to shells in a close neighborhood of $r=0$.
For a black hole to form at the end of collapse, trapped surfaces must appear before the singularity in such a way that no geodesic from the immediate vicinity of the singularity can propagate outside the apparent horizon. 

Therefore a sufficient condition for the formation of a black hole occurs if in the vicinity of $r=0$ we have
\begin{equation}\label{eq:bh-condition}
    t_{ah}(r) \leq t_0  \; .
    %\text{ for } r>0, \text{ near } r=0\,.
\end{equation}
In fact in this case the apparent horizon traps the region surrounding the singularity before the singularity has formed. 
We may expand the apparent horizon curve in the vicinity of $r=0$ as
\begin{equation}
t_{ah}(r)=\frac{4}{(D-1)m_0^{1/4}}\left(1-\frac{m_nr^n}{4m_0}-(m_0^{1/4}r)^\frac{D-1}{D-5}+...\right)\,,
\end{equation}
%\kos{I don't agree with that. What I find if I expand $t_{ah}$ around $r=0$ is}
%\textcolor{red}{\begin{equation}
%t_{ah}(r)=\frac{4}{(D-1)m_0^{1/4}}\left(1-\frac{m_nr^n}{4m_0}(1+\frac{4(m_0^{1/4}r)^\frac{D-1}{D-5}}{D-5})-(m_0^{1/4}r)^\frac{D-1}{D-5}+...\right)\,,
%\end{equation}}
%\kos{It is the same as yours, but in the first order in $n$ there's an extra term, which makes the comparison more complicated.}
%{\bf (DM: I think they are the same. The additional term goes like $r^{n+\frac{D-1}{D-5}}$ which is higher order of the other two terms, i.e. $r^n$ and $r^\frac{D-1}{D-5}$, irrespective of whichever of the two wins.)}
from which we see that if the first non vanishing term $m_n$ has 
\be
n>\frac{D-1}{D-5} \;, 
\ee
then the sufficient condition for the formation of the black hole is met.
On the other hand if $t_{ah}(r) > t_0$ there exist the possibility that the singularity is visible. In fact even if $t_s(r)$ is in the future of $t_{ah}(r)$ for every $r\neq 0$ there is still the possibility that geodesics originating at $r=0$ at the time $t_0$ (where $t_s(0)=t_{ah}(0)=t_0$) may escape. 
In order to determine the condition under which that may happen we need to consider the trajectory of a radial outgoing null geodesic $t_g(r)$ originating at $t_g(0)=t_0$\footnote{Strictly speaking the Cauchy problem for the geodesic equation can not be defined at the singularity and therefore one would have to consider the limit for $r$ going to zero from the right of the initial condition.} and determine if there exist an interval $r\in(0,r_0]$ for which $t_g(r)<t_{ah}(r)$. 
From the line element \eqref{eq:LTB-metric}, the equation for the trajectory of an outgoing radial null geodesic in the marginally bound case, i.e. $E(r)=1$, is
\be 
\frac{dt_g}{dr}=R'(r,t_g(r)) \;,
\ee 
with $R'=a+ra'$ given by
\be 
R'(r,t)=\frac{1}{4m^{3/4}}\frac{1-(D-1)mt-m'rt}{\left(1-\frac{D-1}{4}m^{1/4}t\right)^{(D-5)/(D-1)}}\;.
\ee 

In order for the geodesic to originate from the singularity the integration constant must be chosen in such a way that $t_g(r)\rightarrow t_0$ for $r$ going to zero. We know that such null geodesics exist for any valid initial condition in a positive neighborhood of $r=0$. Therefore we only need to check the condition for the geodesic to be outside the apparent horizon. In order to do that we consider a trajectory
\be 
t_x(r)=t_0+xr^n\;,
\ee 
and check the values of $x$, if any, for which $t_x$ is below the horizon and above $t_g$. If there exist some value of $x$ for which these two conditions are met, then $t_g$ must be also below $t_{ah}$ and therefore outside the horizon in a right neighborhood of the singularity. %For the sake of simpler calculations let us set $m_0=1$ in the following when convenient.
Let us first look at the apparent horizon. We already said that for $n>(D-1)/(D-5)$ the only possible outcome of collapse is a black hole. So we are left with only two possibilities:
\begin{enumerate}
    \item For $n<(D-1)/(D-5)$ we have that
\be 
t_{ah}(r)=t_0-\frac{m_n}{(D-1)m_0^{5/4}}r^n+...
\ee 
and taking $m_n<0$ (and $m_0=1$) we need to choose $x<-m_n/(D-1)$ in order to have $t_x<t_{ah}$.
   \item For $n=(D-1)/(D-5)$ we have that
\be 
t_{ah}(r)=t_0-\frac{1}{D-1}\left(\frac{m_n}{m_0^{5/4}}+4\right)r^n+...
\ee 
and taking $m_n<-4$ (for $m_0=1$) we need to choose $x<-(m_n+4)/(D-1)$ and $x>0$ in order to have $t_x<t_{ah}$.
\end{enumerate}

Let us now look at the condition for $t_x$ to be above $t_g$. Having assumed that $t_g$ is an increasing function in $r$ it is enough to find $x$ for which 
\be 
t_x'=nxr^{n-1}>R'(r,t_x(r)).
\ee 
After some calculations we find
\be 
R'(r,t_x(r))\simeq \Gamma (x) r^{\frac{4n}{D-1}}\;,
\ee 
with
\be \label{NS}
\Gamma (x)=-\frac{\left(\frac{D-1}{4}m_0^{1/4}x+\frac{D+3}{D-1}M_n\right)}{\left(-\frac{D-1}{4}m_0^{1/4}x-M_n\right)^{\frac{D-5}{D-1}}}\;,
\ee 
where we have defined $M_n=m_n/(4m_0)$.
Therefore $t_x'$ goes to zero as $r^{n-1}$ while $R'(r,t_x(r))$ goes to zero as $r^{4n/(D-1)}$ and we have the following two possibilities:
\begin{enumerate}
    \item If $n-1<4n/(D-1)$ we will have that $t_x'>R'$ near $r=0$ for any value of $\Gamma$ and consequently $x$. This means that in this case, for $m_n<0$ it is always possible to find an outgoing null geodesic originating at the singularity.
    \item If $n-1=4n/(D-1)$ we have that $t_x'$ and $R'$ both go to zero as $r^{n-1}$ and therefore we have a value of $M_n$ separating between the two outcomes and given by the solution of $nx=\Gamma (x)$. Then a naked singularity will form if values of $M_n$ exist such that the condition in Eq.~\eqref{eq:bh-condition} is violated (i.e. $0<x<-(m_n+4)/(D-1)$ with $m_n<-4$) and $nx>\Gamma (x)$.
\end{enumerate}
Notice that $n-1=4n/(D-1)$ is exactly equivalent to $n=(D-1)/(D-5)$ and therefore this condition is consistent with the one found previously.

\begin{figure}[!ht]
    \centering
    \includegraphics[scale=0.45]{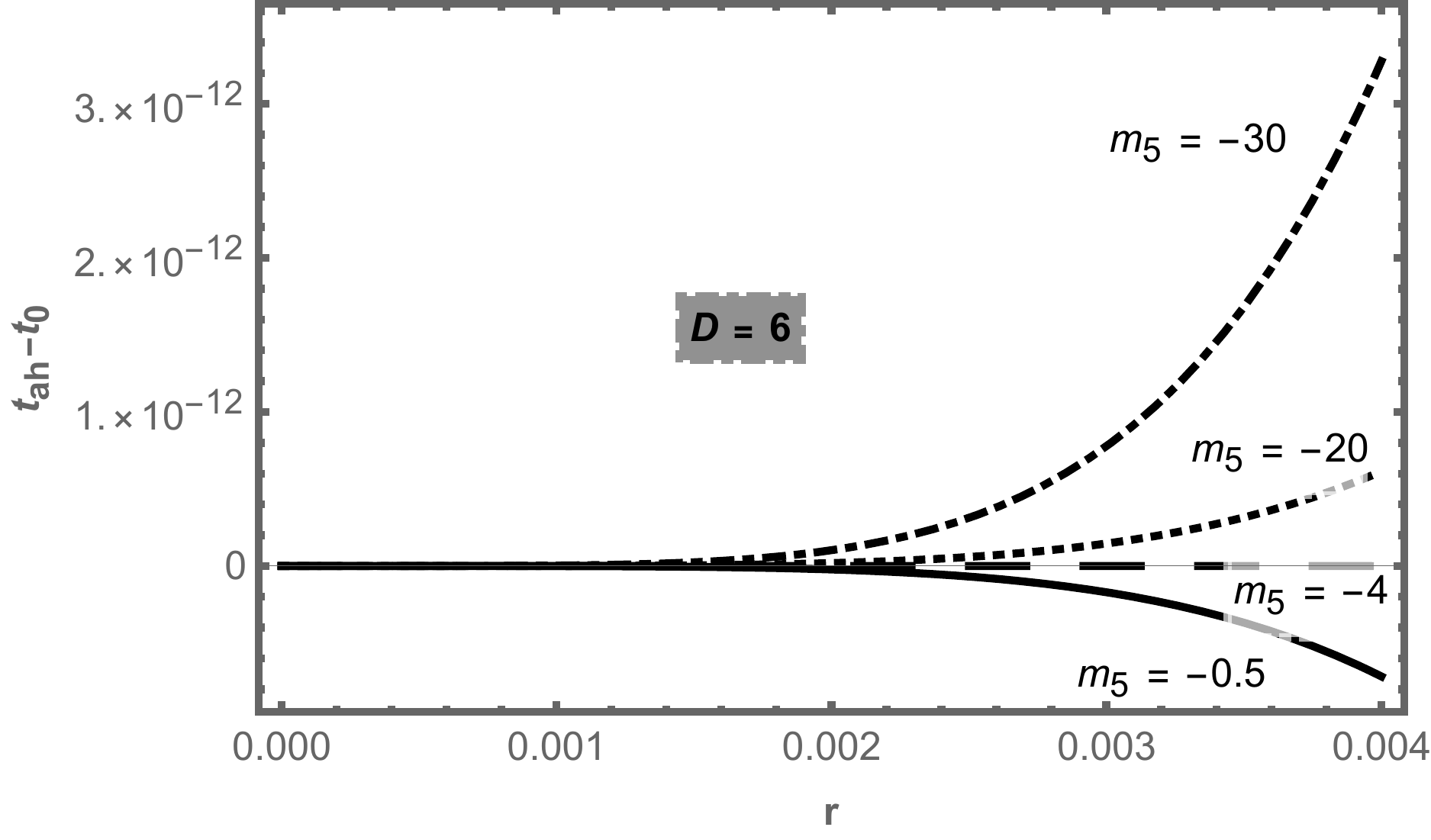}
    \includegraphics[scale=0.45]{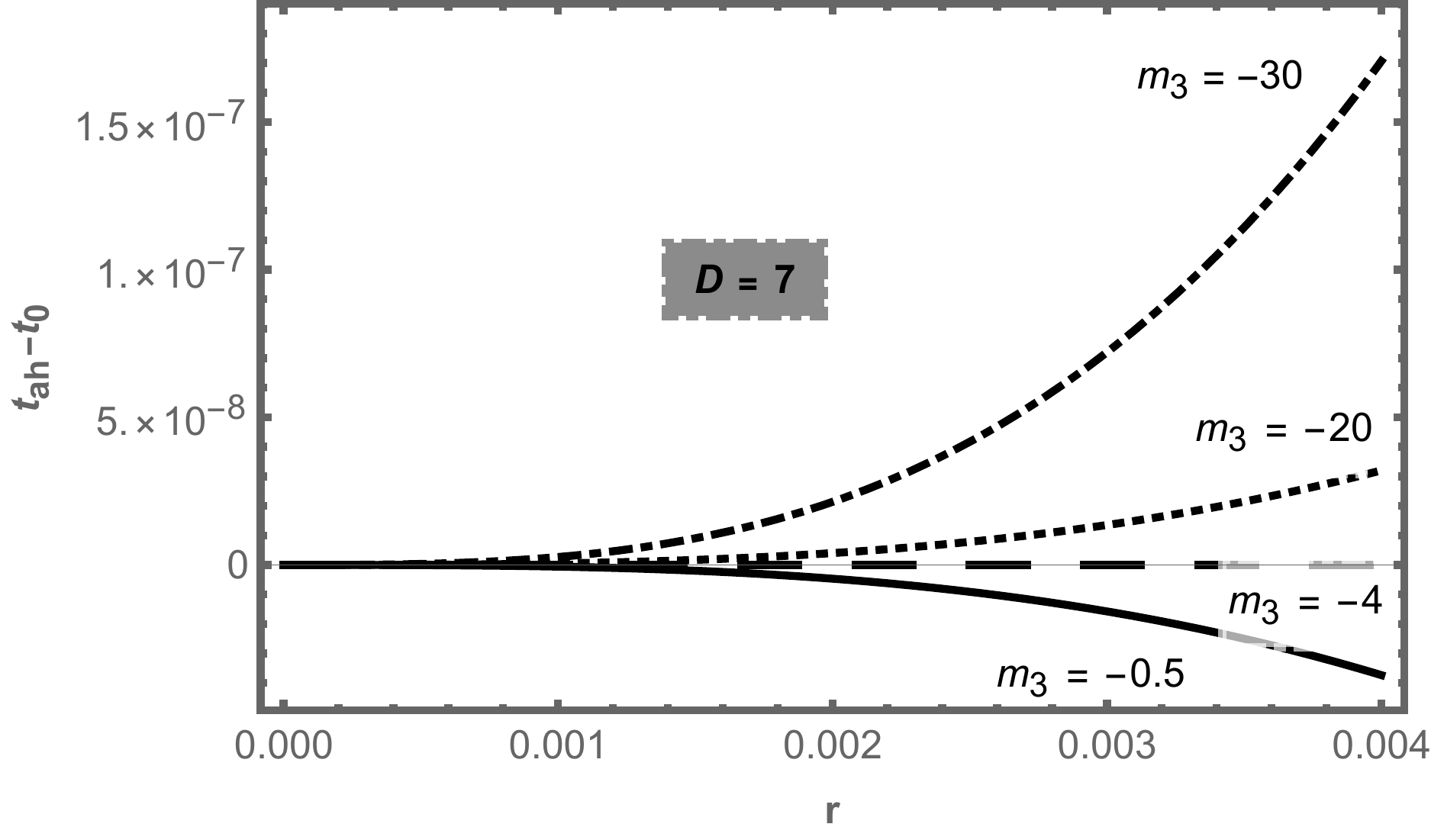}
    \caption{Left panel: The difference $t_{ah}-t_s(0)$ in $D = 6$ is plotted for different values of the mass parameter $m_5$. When this condition is non-positive, the formation of black hole occurs. Otherwise, we can have the formation of a naked singularity. Right panel: The difference $t_{ah}-t_s(0)$ in $D = 7$ for different values of the mass parameter $m_3$. When this condition is non-positive, the formation of black hole always occurs. Otherwise, we can have the formation of a naked singularity. In both panels we have set $m_0 = 1$.}
    \label{fig:Black-hole-condition-D=7}
\end{figure}

%Let us consider $D = 6$. 
Let us now look at some specific cases and for simplicity let us take $m_0=1$ (this is always possible since $m_0$ only sets the scale of the radial coordinate $r$) when looking for the conditions for the formation of naked singularities.
For the $D = 6$ case we have that $(D-1)/(D-5)=5$ and therefore the sufficient condition for the formation of a black hole is violated for the first four terms, i.e. $n = 1,2,3,4$ when $m_n<0$. For $n = 5$ we can have a black hole only if
\begin{equation}
    \frac{4}{5} \left(-\left(r^5 \left(m_0+m_5 r^5\right)\right)+\frac{1}{(m_0+m_5 r^5)^{1/4}}-\frac{1}{m_0^{1/4}}\right) \leq 0 \,, \quad \text{near } r = 0\,.
\end{equation}
In the left panel of Fig.~\ref{fig:Black-hole-condition-D=7} we can see that, setting $m_0 = 1$, we can have a black hole when $0 \geq m_5 \geq -4$. On the other hand, for $n\geq 6 $, the condition in Eq.~\eqref{eq:bh-condition} is always satisfied and a black hole always forms. 
%\begin{figure}[!ht]
%    \centering
%    \includegraphics[scale=0.5]{BH-formation-D=6-2.pdf}
%    \caption{The difference $t_{ah}-t_s(0)$ in $D = 6$ is plotted for different values of the mass parameter $m_5$. When this condition is non-positive, the formation of black hole occurs. Otherwise, we have the formation of a naked singularity. We have set $m_0 = 1$.}
%    \label{fig:Black-hole-condition-D=6}
%\end{figure}
For $n=5$ and $m_5<-4$ we need to also look at the condition $nx>\Gamma (x)$ from Eq.~\eqref{NS}. The condition $nx=\Gamma (x)$ can be rewritten as
\be
(5x+9m_5)^5-\frac{5}{4}(20x)^5(x+m_5)=0\;,
\ee 
which is a well defined algebraic equation for which it is possible to find values of $m_5<-4$ such that it has solutions for $0<x<-(m_5+4)/5$. As mentioned before, a naked singularity will then be final outcome of collapse for those values $m_5$ for which we can satisfy $nx>\Gamma (x)$ with $0<x<-(m_5+4)/5$.

In $D = 7$, as expected from the previous section, we have a case qualitatively equivalent to GR, and thus the condition in Eq.~\eqref{eq:bh-condition} is violated for $n = 1,2$ while for $n= 3$ we have a black hole as final state if $m_3 \geq - 4$ (with $m_0 = 1$), as we can see from the right panel of Fig.~\ref{fig:Black-hole-condition-D=7}.
For the rest of the modes, i.e. $n \geq 4$ the condition \eqref{eq:bh-condition} is always satisfied and collapse always results in a black hole.
The condition $nx=\Gamma (x)$ can be rewritten as
\be 
\frac{1}{16}(6x+5m_3)^3-27(2x+m_3)x^3=0\;,
\ee 
which is again an algebraic equation for which it is possible to find values of $m_3<-4$ such that it has solutions for $0<x<-(m_3+4)/6$.

\begin{figure}[!ht]
    \centering
    \includegraphics[scale=0.43]{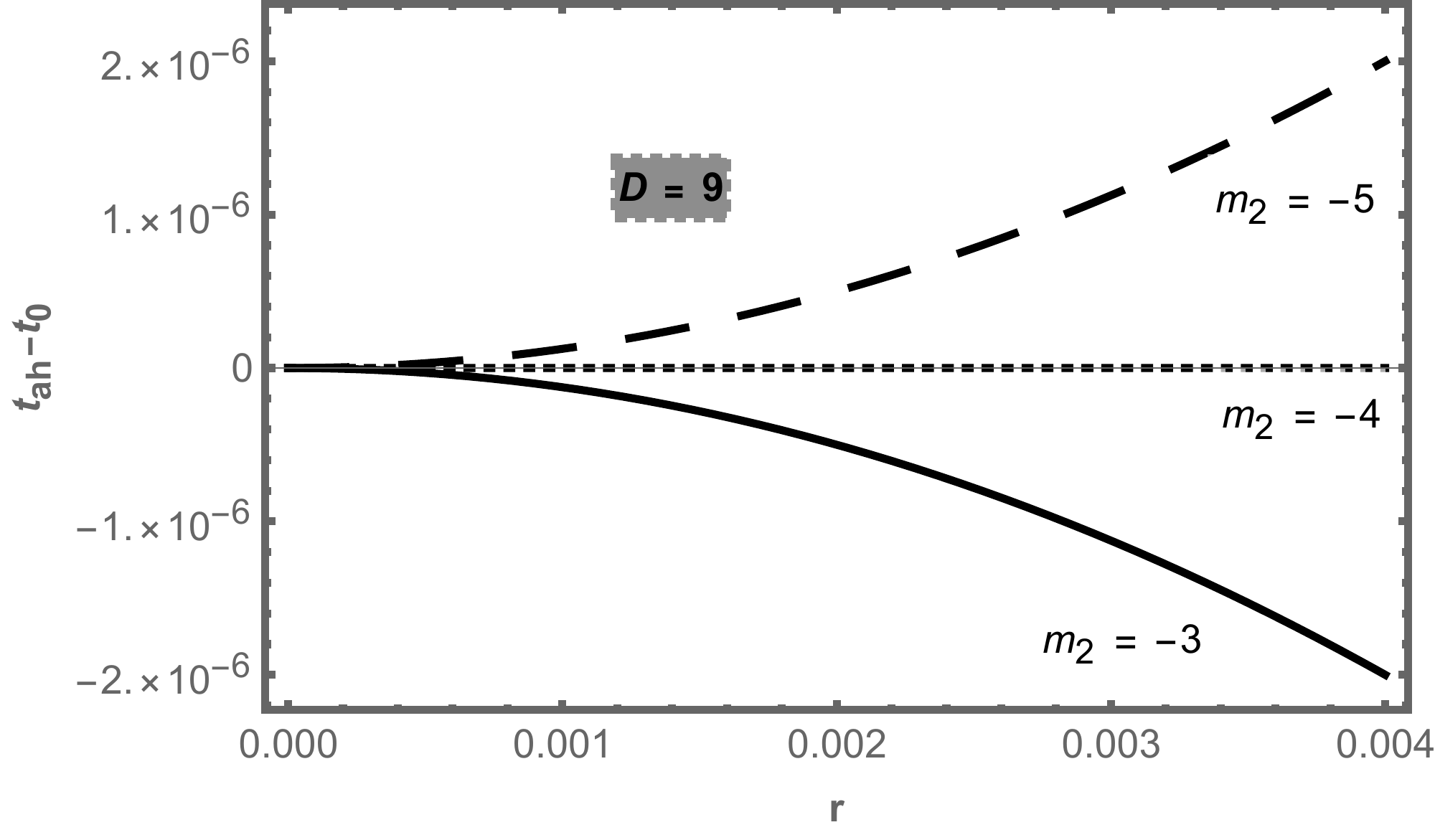}\    \includegraphics[scale=0.43]{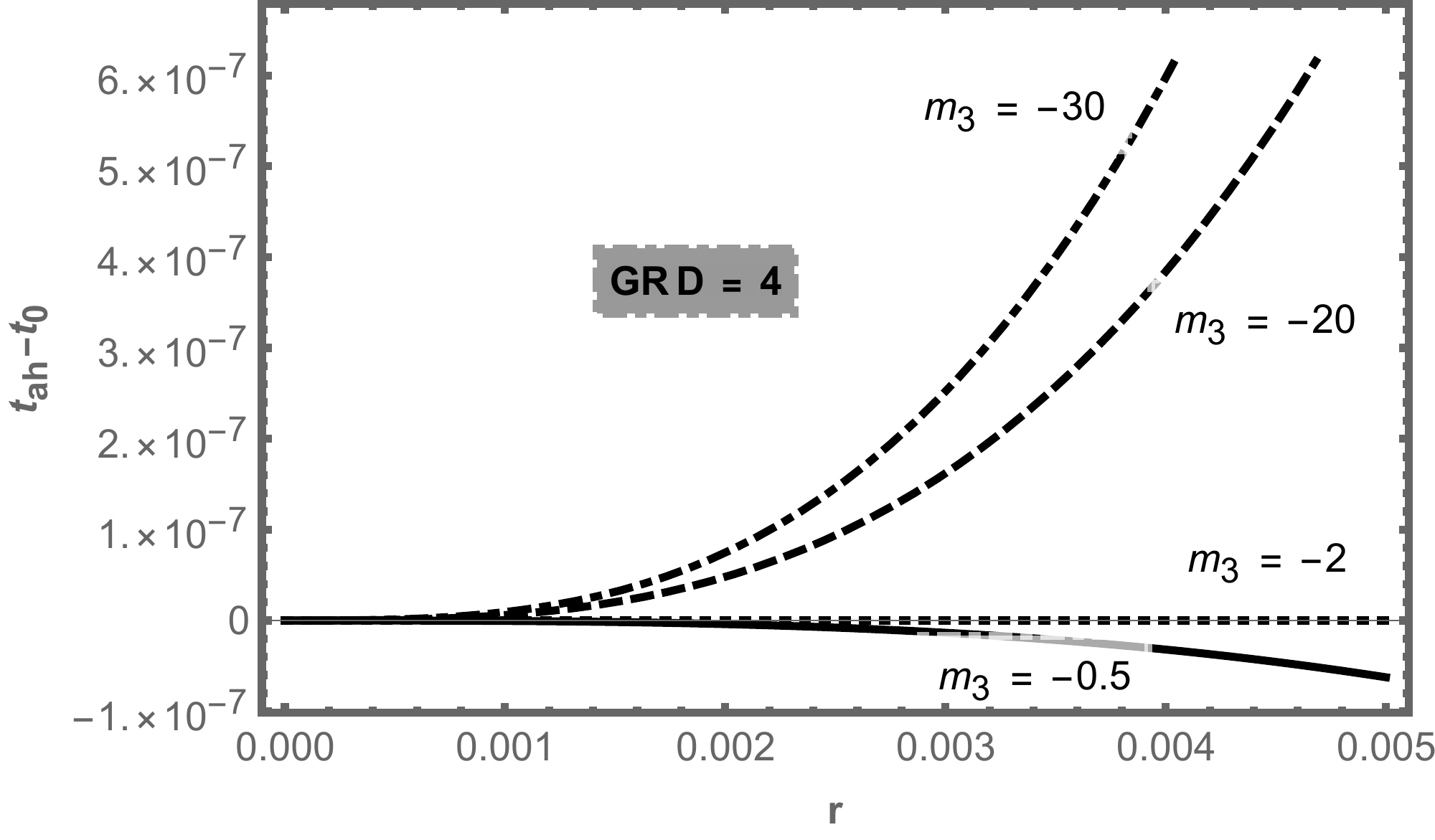}
    \caption{The difference $t_{ah}-t_s(0)$ in $D = 9$ pure Gauss-Bonnet (left panel) and $D = 4$ GR (right panel panel) is plotted for different values of the mass parameter $m_2$ and $m_3$ respectively. These are the critical modes for which we can get both outcomes. 
    %when $m_3 \gtrsim -2$ and $m_2 \gtrsim -4$ we get black holes, while when $m_2 \prec -0.5$ and $m_1 \prec 0.01$ we get naked singularities. 
    In both panels we have set $m_0 = 1$.}
    \label{fig:Black-hole-condition-D=8-9}
\end{figure}

As explained in the previous section, $D = 6$ pure Gauss-Bonnet gravity is weaker compared to GR. In particular, the gravitational potential falls as $1/r^{(D-5)/2}$. 
%{\bf (DM: I guess it should be "the gravitational potential falls as $1/r^{(D-5)/2}$", right?)}
That is why, the critical mode is $m_5$ for $D=6$, while in $D=7$ and in GR it is $m_3$. As expected in $D > 7$ the gravitational force for pure GB theory becomes stronger with respect to GR and thus the value of $n$ for the critical mode will decrease. Indeed in $D = 8$ pure GB gravity there is no critical mode because $(D-1)/(D-5)=7/3>2$. Then for $m_1<0$ and $m_2<0$ it is always possible to find null geodesics escaping from the singularity while for $n\geq 3$ the black hole is the only possible outcome.
%the mode for which we get both black holes and naked singularities is the second one, $m_2$, while 
Finally, as we show in the left panel of Fig.~\ref{fig:Black-hole-condition-D=8-9}, in $D = 9$ we get that the critical mode is $m_2$. The condition $nx=\Gamma (x)$ can be rewritten as
\be 
\left(2x+\frac{3}{4}m_2\right)^2-2x^2(4x+m_2)=0\;,
\ee 
which is a cubic algebraic equation. Again for $m_1<0$ there always exist null geodesic escaping the singularity and again it is possible to find values of $m_2$ for which $nx>\Gamma (x)$ with $0<x<-(m_2+4)/8$ therefore leading to the formation of a naked singularity. For $n\geq 3$ the black hole is the only possible outcome.

 Since in $D=10$ we have $n=9/5<2$ and since $(D-1)/(D-5)\rightarrow 1$ as $D$ grows we see that for all $D>9$ the only mode that may lead to the formation of a naked singularity is $m_1$. However, one could argue that $m_1 \neq 0$ gives rise to unphysical density profiles with a cusp at $r=0$ and therefore $m_1$ should be set to zero. Thus for $D>9$ the black hole is the only possible final outcome.
%, however, our equations are not affected by any assumptions on $m_n$.

For comparison we may look at what happens in higher dimensional collapse in GR \cite{2002PhRvD..65j1501J, Goswami:2002he, Goswami:2004gy, Goswami:2006ph}. Similarly to what happens in pure GB, gravity becomes stronger in higher dimensions and therefore the chance of black hole final outcomes increases. In GR with $D=4$ for the first two modes, i.e. for $n = 1,2$, the condition \eqref{eq:bh-condition} is violated and we may have a naked singularity for negative values of $m_1$ and $m_2$. Similarly to what was discussed before in the pure GB case with $D=7$ the mode $m_3$ is the one separating the two possible outcomes while for all higher modes the black hole is the only possible final state, as illustrated in the right panel of Fig.~\ref{fig:Black-hole-condition-D=8-9}. In GR black holes are the only possible outcome in $D>6$, while in $D=6$ and $D=5$ the mass mode separating the two possible outcomes is $m_1$ and $m_2$ respectively. The comparison is summarized in Table~\ref{tab:recap2}.
%This means that the strong gravity regions can communicate with outside observers and thus we have a naked singularity. 

\begin{table}[ht]
\begin{tabular}{|c|c|c|c|c|c|c|}%
\hline
  & $D = 4$ & $D = 5$ & $D = 6$ & $D = 7$ & $D = 8$ & $D = 9$\\
\hline
pure Gauss-Bonnet & - & - & $n=5$ & $n=3$ & $n=7/3$ & $n=2$   \\
\hline
General Relativity & $n=3$ & $ n=2$ & $n=1$ & - & - & -   \\
\hline
\end{tabular}
\caption{The critical mass modes for which both naked singularities and black holes can be formed in the inhomogeneous dust collapse are summarized in pure Gauss-Bonnet gravity and GR. For the lower modes naked singularities can occur, while for the higher ones only black holes are formed. The cases $D = 4, 5$ in pure Gauss-Bonnet are excluded, while in $D>6$ in GR and $D>9$ GB we have no naked singularities.}\label{tab:recap2}
\end{table}

\section{Conclusions}\label{conc}
 
Pure Lovelock universalizes the kinematic property of gravity in the critical odd $D=2N+1$ dimensions. Note that 
Einstein gravity is pure Lovelock $N=1$ and it is kinematic in $D = 2\times 1+1=3$. Kinematicity means Riemann curvature is 
entirely given in terms of Ricci and hence there can exist no non-trivial vacuum solution. It has been shown that Lovelock Riemann
tensor which has been defined in \cite{Camanho:2015hea} is in fact given in terms of the corresponding Ricci in all $D=2N+1$ 
dimensions, and so whenever the latter vanishes the former vanishes. Another distinguishing property of it is the existence of 
bound orbits around a static object in dimensions $2N+1<D<4N+1$ which implies that for $N=1,$ GR bound orbits can exist only in $D=4$. Thus bound orbits in higher dimensions can exist only in pure Lovelock gravity which provides a case for the study of such theories. 

In pure Lovelock gravity, the gravitational potential goes as $1/r^\alpha$ where $\alpha=(D-2N-1)/N$ which is $\leq 1$ for $2N+1<D\leq3N+1$ while it is otherwise for $D>3N+1$. It is therefore expected that collapse would proceed slower for $2N+2\leq D<3N+1$ with respect to the case when the potential goes as $1/r$. In the critical dimension, $D=2N+1$, collapse proceeds uniformly with $\dot a = const.$ as gravity is kinematic there and hence no acceleration. 
%On the other hand the presence of non vanishing spatial curvature may accelerate (in the case $k>0$) or decelerate (in the case $k<0$) collapse. 
Note that potential goes as $1/r$ \cite{Chakraborty:2016qbw} for all $D=3N+1$; i.e., for Einstein in 4D and pure GB in 7D, gravitational dynamics should be indistinguishable. 
%This is however so only for the case $k=0$ while it differs when $k\neq 0$. This is because the curvature of the 3-space geometry, when positively curved, 'helps' collapse while negative curvature works otherwise. Intuitively, it is natural to expect that positive spatial curvature favours collapse, while negative one opposes it. 

%What we had set out to study in this paper was to demonstrate in the context of gravitational collapse process the effect of gravity becoming  weaker than GR in $2N+2\leq D <3N+1$ and stronger outside of this window. What is then expected is that collapse should proceed slower than GR in this window and faster outside it. Secondly we believe that the effect of spatial curvature $k$ on collapse is being explored for the first time for pure GB gravity. As noted before, positive $k$ goes along collapse while negative against it. Through various examples we have compared the evolution of various quantities of interest in different dimensions and also relative to GR. All this has been demonstrated through figures and tables.   

Furthermore we considered marginally bound inhomogeneous dust collapse in pure GB gravity and derived the condition for the formation of naked singularities, similarly to what had been done in GR in $D=4$ \cite{Joshi:1993zg} and in higher dimension \cite{Goswami:2004gy}. We calculated the first non vanishing mass mode in the expansion of the energy density which separates the case where naked singularities can for from the case leading only to black holes as final states of collapse. We showed that, consistently with what was mentioned earlier, pure GB in $D=7$ and GR in $D+4$ exhibit the same behavior. Also we showed that in $D>9$ pure GB collapse leads only to the formation of black holes.

%\begin{table}[!ht]
%\begin{tabular}{|c|c|c|c|c|}%
%\hline
% & General Relativity & \multicolumn{3}{c|}{Pure Gauss-Bonnet}  \\
%\hline\hline
%Spatial Curvature & $D = 4$ & $D = 6$ & $D = 7$ & $D = 8$ \\
%\hline
%$k=1$ &  &  &  &   \\
%\hline
%$k=0$ &  &  &   &  \\
%\hline
%$k=-1$ &   &  &  & \\
%\hline
%\end{tabular}
%\caption{Homogeneous dust collapse in pure Gauss-Bonnet gravity in $D = 6, 7$ and $8$ dimensions compared to $D = 4$ Oppenheimer-Snyder-Datt collapse in general relativity, for the three different spatial curvatures, $k = -1,0,1$. }\label{tab:recap}
%\end{table}

%{\bf (DM: I get something different. Just checking for the sufficient condition for black hole formation I get that if the first non vanishing term $m_n$ has $n>(D-1)/(D-5)$ then a BH forms. This gives me that there is no critical $m_n$ for $D=8$, because $n>7/3$ and therefore for $m_2<0$ we have naked singularities while for $m_3$ we always have black holes, and $m_2$ for $D=9$ (because $n=2$.)}\kos{I wrote some comments above.}

\begin{acknowledgements}
The work was supported by Nazarbayev University Faculty Development Competitive Research Grant No. 11022021FD2926 and by the Hellenic Foundation for Research and Innovation (H.F.R.I.) under the ``First Call for H.F.R.I. Research Projects to support Faculty members and Researchers and the procurement of high-cost research equipment grant'' (Project Number: 2251). The work was supported by the PNRR-III-C9-2022–I9 call, with project number 760016/27.01.2023 and is based upon work from COST Action CA21136 Addressing observational tensions in cosmology with systematics and fundamental physics (CosmoVerse) supported by COST (European Cooperation in Science and Technology). ND wishes to thank Albert Einstein Institute, Golm for a summer visit.   
\end{acknowledgements}

\bibliographystyle{utphys}
\bibliography{references}

\end{document}